**Prediction of nanobubble-assisted focused ultrasound-induced blood-brain barrier opening with machine learning**

Wenjing Li[1], Chenchen Bing[3], Haixin Dai[1], Rajiv Chopra[3,†], Qian Wang[2], Bingbing Cheng[1]*

Affiliations

1. Translational Research in Ultrasound Theranostics Laboratory, School of Biomedical Engineering, ShanghaiTech University, Shanghai, China

2. School of Biomedical Engineering, ShanghaiTech University, Shanghai, China

3. Department of Radiology, UT Southwestern Medical Center, Dallas, TX, USA

[†]: The author has moved to 8Fold Mfg LLC.

*Corresponding author: Bingbing Cheng, Translational Research in Ultrasound Theranostics Laboratory, School of Biomedical Engineering, ShanghaiTech University, 393 Middle Huaxia Road, Shanghai, 201210, China; Email: chengbb@shanghaitech.edu.cn

**Abstract**

Novel approaches for predicting the outcomes of blood-brain barrier (BBB) opening with focused ultrasound (FUS) and microbubbles are highly desired. This study aims to explore machine learning-based methods for reliably predicting the FUS-induced BBB opening efficacy and safety. **Methods**: Sixteen female rats were used in this study. An acoustic feedback-controlled FUS system ($f_0$: 0.5MHz) was used for the BBB opening with the infusion of custom-made nanobubbles/Definity. Evans Blue was injected for the BBB opening efficacy verification and the brain tissue was harvested for the safety assessment. Acoustic emissions were recorded, preprocessed and fed into three machine learning models for BBB opening outcomes prediction. Conventional stable and inertial cavitation dose were also calculated. **Results**: Among the tested machine learning models, a modified Support Vector Data Description (mSVDD) model achieved the best performance in the BBB opening efficacy and safety prediction with an accuracy of 85.0 ± 16.6% and 62.5 ± 12.8%, respectively. Conventional stable and inertial cavitation dose-based prediction has a prediction accuracy of 80.0% in efficacy and 34.3% in safety, respectively. The mSVDD model trained with the overall bubble response (0-2 MHz) performed better than that trained with the ultra-harmonic bubble response (0.7-0.8 MHz) in both efficacy prediction (85.0 ± 16.6% vs 76.0 ± 8.0%, p=0.04) and safety prediction (62.5 ± 12.8% vs 55.0 ± 10.7%, p>0.05). It is also found that the mSVDD model trained with nanobubble data cannot be directly applied to Definity. **Conclusion**: Our investigations demonstrated that it is feasible to achieve a reliable prediction of FUS-induced BBB opening outcomes with machine learning and acoustic signals from stimulated nanobubbles. This study provided a new approach for the prediction of FUS-BBB opening outcomes with a clinical translation potential.

**Introduction**

The blood-brain barrier (BBB) is a semipermeable physiological structure that protects the brain from harmful substances circulating in the bloodstream [1]. With the tight junctions between endothelial cells and various efflux proteins, such as P-glycoproteins (P-gp), the BBB can maintain the brain homeostasis [2]. However, it also presents as a bottleneck for the efficient delivery of therapeutics into the brain to treat brain diseases. Focused ultrasound (FUS) combined with microbubbles/nanobubbles can locally, non-invasively and temporarily open the BBB [3]. A variety of therapeutics have been successfully delivered to the brain via this method in both pre-clinical and clinical studies, such as: Temozolomide, Doxorubicin, antibodies, viral vectors for gene therapy, and drug conjugated nanoparticles [4]. Multiple ultrasound contrast agents have been used in this application, including commercial agents: Definity, Optison, SonoVue, and custom-made nanobubbles [5-8]. Noticeably, in some pre-clinical studies, nanobubbles have shown favorable results in FUS-induced BBB opening due to their smaller size, longer in-vivo circulation time, longer in-vitro persistence time, and stable harmonic emissions during acoustic feedback-controlled treatments [8, 9]. Our previous study demonstrated that two hyperpolarized $^{13}$C-labelled substrates ([1-$^{13}$C]pyruvate and [1-$^{13}$C]glycerate) were efficiently delivered to the rat brain with FUS and nanobubbles, providing enhanced sensitivity to image brain metabolism with hyperpolarized $^{13}$C MR imaging [10].

The outcomes of FUS-induced BBB opening include efficacy and safety. 1) For efficacy, the current gold standard to verify BBB opening is contrast-enhanced T1-weighted magnetic resonance imaging (T1w MRI) [11-14]. It can indicate the BBB opening and closure by measuring the contrast enhancement in T1w MR images. However, in clinical settings, physicians do not perform this procedure for every patient due to the requirement of contrast agent injection, the long scanning time, and an MRI scanner. In pre-clinical studies, BBB opening efficacy is often assessed by the dye leakage at the FUS targets, such as Evans Blue. While reliable, it is not convenient due to the complicated blood perfusion and brain harvesting procedures [8, 15] and it can only be applied in acute studies. 2) For the safety consideration, there is evidence showing that FUS-induced BBB opening might cause tissue damage, including: red blood cell (RBC) extravasation or hemorrhage [4, 16]. The safety was usually assessed by T2-/T2*-weighted MRI or histology evaluation in clinical and pre-clinical studies. However, these verifications are time consuming, complicated, and limited by the access of MRI scanner.

In recent years, researchers have been investigating using the acoustic signals generated from ultrasound-stimulated bubble agents as real-time feedback to control the ultrasound energy during the treatment [17]. Multiple controlling algorithms have been developed based the harmonic, sub-harmonic, or ultra-harmonic emissions. Moreover, the acoustic signals detected during FUS exposures have also been used as indicators for the treatment outcomes, as their spectral characteristics uncover features of the underlying bubble oscillation dynamics. Conventionally, stable cavitation and inertial cavitation are used as indicators for predicting opening efficacy and tissue damage, respectively. Stable cavitation, where the microbubbles repeat a process of volumetric oscillation often results in a shear force to break the local tight junctions to open the BBB [18, 19]. Inertial cavitation, on the other hand, usually detected as broadband acoustic emissions, indicates potential tissue damage caused by bubble collapse and strong mechanical stress [16]. Numerous reports have been attempted to correlate the acoustic emissions (such as: stable cavitation dose, inertial cavitation dose) to the BBB opening outcomes, including: the magnitude of increased BBB permeability, the BBB opening dynamics, and the presence of damage. For instance, Xu et al. reported that the inertial cavitation dose has a linear correlation with the scale of histologic-level tissue damage in mice [20]. However, several studies have suggested the detection of broadband emissions does not always result in damage, whereas in other cases RBC extravasations have been observed in the target without detecting any significant wideband emission [17].

Machine learning has gained a lot of interest in the ultrasound field in the past few years [21]. It has not only been used in diagnostic ultrasound such as image reconstruction, segmentation, and beamforming; but also in therapeutic ultrasound for treatment outcome prediction and sophisticated patient selection. Several articles showed preliminary results of combining machine learning models with multi-parametric MRI in treatment of uterine fibroids using MR-guided high-intensity focused ultrasound (MR-HIFU) ablation to improve the clinical outcome [22, 23].

In this study, we sought to investigate machine learning-based methods for fast prediction of the treatment outcomes of FUS-induced BBB opening with nanobubbles in terms of opening efficacy and safety, based on the acoustic signals collected during FUS exposures. To achieve this goal, we developed a prediction algorithm based on three machine learning classifiers, including: modified Support Vector Data Description (mSVDD), isolation Forest (iForest), and Minimum Covariance Determinant (MCD). We then investigated the training methods to result in the best prediction performance. Upon comparing the testing results, we identified the most suitable machine learning classifier for this study. To evaluate the robustness and compatibility of the algorithm, we further examined whether a machine learning model trained with one bubble type can be applied directly across different bubble formulations.

**Methods**

Real-time feedback-controlled focused ultrasound system

The acoustic feedback-controlled FUS system is similar to our previous report [9] (Figure 1A). The FUS transducer has a fundamental frequency ($f_0$) of 0.5 MHz with a 26-mm diameter aperture. A hydrophone with a center frequency of 0.75 MHz was inserted into the transducer to monitor and collect the acoustic emissions from stimulated bubble agents. The transducer was connected to a custom-built driving system and then mounted on a stereotaxic apparatus. The acoustic signals were digitized with a data acquisition card (DAQ, sampling rate: 20 MHz, ATS460, Alazar Technologies Inc, Pointe-Claire, QC, Canada). The Fast Fourier Transform (FFT) was performed on the collected acoustic signals (Figure 1B) and the area under the curve (AUC) at the ultra-harmonic frequency (1.5*$f_0$, 0.75 MHz) within a 100 kHz bandwidth was calculated for each ultrasound burst. During treatment, the ultrasound pressure was adjusted in real time based on the AUC of current acoustic signals with a step size of 0.01-0.03 MPa.

Animal studies

*Animal preparation*

Female rats (Sprague Dawley, 230-300 g, n=16) were used in this study. All procedures were approved by UT Southwestern Institutional Animal Care and Use Committee and followed guidelines set forth by the Guide for the Care and Use of Laboratory Animals. The animals were administered anesthesia by inhaling 2-3% isoflurane and 1-2 L/min of 100% oxygen. A 24-gauge intravenous catheter was inserted into the lateral tail vein for bubble administration, while vital signs were continuously monitored using a pulse oximeter. The body temperature was maintained at 36.5°C with a physiologic monitoring system (PhysioSuite, Kent Scientific Corp., Torrington, CT, USA). The hair over the animal's head was removed for acoustic coupling. After preparation, the animal was then transferred onto a stereotaxic apparatus and stabilized using ear bars and a bite bar. During sonications, a custom-built nose cone was utilized to deliver inhalant anesthetic to the animal.

A small incision was made over the skull to identify cranial landmarks for atlas registration. Before ultrasound exposure, Evans blue (2%, 3.33 ml/kg dosage) was injected via the tail vein catheter and allowed to circulate for a minimum of 3-5 minutes for the BBB opening verification. Two bubble agents were used in this study: a custom-made nanobubble and a commercially available microbubble Definity (Lantheus Medical Imaging, Billerica, MA, USA). The nanobubble (mean diameter: 313±13.2 nm) was made according to the method described in our previous report [8]. Three rats (6 targets in total) were treated with Definity and the other thirteen (35 targets in total) were treated with nanobubbles. Bubbles were administered through the tail vein catheter via infusion at 0.3 ml/min infusion rate according to the sonication protocol.

Animals were euthanized using transcardiac perfusion with saline and 10% buffered formalin approximately 5-10 minutes after the treatment. The brain was harvested immediately and fixed in formalin for further evaluation.

*Ultrasound exposure*

The bubble solutions of nanobubbles and Definity were prepared to reach a total gas volume of 1.1-1.2 μl/ml. The volume of a single bubble could be calculated using equation:

$$V_{bubble} = 4/3\ \pi(d/2)^3 \times 10^{-9}$$

where $d$ is the mean diameter of the bubbles. The dosage of each bubble formation could be then derived based on the total gas volume and bubble volume. At each target, ultrasound exposures were delivered with the 10 ms pulse length, 1 Hz pulse repetition frequency, and total duration of 100 seconds. During the treatment, the ultrasound pressure was adjusted by the controlling algorithm mentioned above. Based on prior characterization results and considerations of signal-to-noise ratio, the efficacy threshold (AUC) for controlling BBB opening, was set at 5000, while the corresponding pressure in the system was approximately 0.3-0.4 MPa.

Prediction algorithm

The workflow of the prediction algorithm is shown in Figure 2. It is consisted of three parts: data preprocessing, dataset partition, and model training and evaluation.

*Data preprocessing*

Each sample of the acoustic data in the time domain contained 100×200,000 data points, corresponding to 100 bursts. The FFT was performed on the time-domain acoustic signal to get the frequency spectra, followed by normalization. Then, data in two frequency bands (0.7-0.8 MHz and 0-2 MHz) were selected for the following data processing. The former frequency band corresponds to the ultra-harmonics of the bubble emission, while the later corresponds to the overall response of stimulated bubbles.

*Dataset partition*

In the prediction of the BBB opening efficacy, a positive label indicates that the BBB was open, while a negative label indicates that the BBB was not open. In the prediction of safety, a positive label represents that no hemorrhage/micro-hemorrhage present, while a negative label indicates the presence of hemorrhage/micro-hemorrhage. For each task, the training dataset was generated with 90% of the positive data randomly selected. The testing dataset was comprised of the negative data and the remaining positive data. This process was repeated twenty times to reduce the potential impact of random dataset partitioning on the results.

*Machine learning-based prediction models*

Three machine learning-based classifiers were used in this study, including: a modified Support Vector Data Description (mSVDD), isolation Forest (iForest), and Minimum Covariance Determinant (MCD).

*mSVDD model*

The mSVDD model [24, 25] obtains a hyper-sphere boundary with minimal volume around a dataset, such that the normal data are inside the boundary and the outliers are outside. It is based on the Support Vector Machine (SVM) algorithm, as a hyper-sphere with center $\boldsymbol{a}$ and radius $R$ that contained as many as possible the training dataset is fitted by the model. In the event that the distance calculated by a given test sample exceeded the radius of the center of the hypersphere obtained via the fitted mSVDD, the sample was deemed a negative outlier. Conversely, if the distance calculated by the test sample was less than or equal to the radius of the center of the hypersphere, the sample was classified as positive. To determine the hypersphere, we adjusted the tightness of the boundary automatically and optimized though a method proposed by Xiao, Y., et al. [26]. The trained mSVDD model was utilized for evaluation on the testing set.

*iForest model*

The iForest model [27, 28] is an anomaly detection algorithm that uses decision trees to isolate anomalies as they have shorter average path lengths to the root of the tree than the normal data points. In this study, the number of trees in the forest was n = 1000, a value larger than the default number due to multiple features [29]. For improved detection performance without redundant features, 80% of the features were used to train each base estimator. The testing sample was evaluated in the model, and the average of the division depth across all decision trees in the forest was computed. Samples that have a smaller depth, indicating less distinguishability, were classified as negative, while those with a larger depth were classified as positive.

*MCD model*

The MCD model [30] is one of the first affine equivariant and highly robust estimators of multivariate location and scatter, which finds the subset of the data with the smallest determinant of the covariance matrix. In this study, the principle component analysis (PCA) algorithm was used to reduce the dimensionality of the training set while retaining 99% of the original data variance, and then the covariance matrix was computed using MCD model. For the testing dataset, the cardinality distribution of the classical Mahalanobis distance (MD) was calculated (the statistical distribution of MD was based on the cardinality distribution) to get the threshold. Then the confidence level was set to 0.8 and the degree of freedom was set to the number of dimensions left after dimensionality reduction. Subsequently, the MD of the testing sample was calculated based on the matrix derived from the training set. This distance was then compared against the pre-defined threshold value, whereby if the MD exceeded the threshold, the sample was deemed to be a negative outlier, and if the MD is below the threshold, the sample was classified as positive.

The mSVDD, iForest and MCD models were generated in Python with classical SVDD from KePeng Qiu [31], and the last two from the Scikit-learn library to define the two classifiers and their associated hyperparameters.

The confusion matrix used in this study is shown in Table 1. TP is true positive, meaning the number of correctly identified positive data; TN is true negative, meaning the number of correctly identified negative data; FP is false positive, meaning the number of incorrectly identified positive data, namely the non-detected negative data; FN is false negative, meaning the number of incorrectly identified negative data, namely the non-detected positive data.

To evaluate the performance of the machine learning based algorithms, the following metrics were calculated, including accuracy (ACC), recall, and precision.

$$ACC = (TP+TN)/(TP+TN+FP+TN)$$

$$Recall = TP/(TP+FN)$$

$$Precision = TP/(TP+FP)$$

The F1-Score reflected the average of precision and recall, which was calculated as:

$$F1\text{-}Score=(2*Recall*Precision)/(Recall+Precision)$$

The Matthews correlation coefficient (MCC) is a metric that ranges from -1 to +1, where +1 indicates prefect prediction, 0 represents random prediction, and -1 indicates complete inconsistency between the prediction and observation. Considering the imbalanced dataset in this study, the MCC was obtained and compared among different prediction models [32].

MCC was calculated by the following equation:

$$MCC=(TP*TN-FP*FN)/\sqrt{((TP+FN)(TP+FP)(TN+FP)(TN+FN))}$$

Conventional cavitation-based methods for BBB opening outcome prediction

The stable cavitation dose (*SCD*) and inertial cavitation dose (*ICD*) were calculated according to a previously reported method [33, 34]. Acoustic signal detected in each ultrasound pulse was transformed to the frequency domain using FFT. For stable cavitation, considering the center frequency of the hydrophone, the emission signals in the frequency band of 750 ± 25 kHz (ultra-harmonic) was defined as the stable cavitation dose for each pulse (*$SCD_p$*). For inertial cavitation, the inertial cavitation dose for each pulse (*$ICD_p$*) was calculated as the root mean square amplitude of the frequency spectra in the 0.6 to 2 MHz range, which excludes the harmonic frequencies with 360 kHz bandwidth and sub-/ultra-harmonic frequencies with 100 kHz bandwidth. Through the treatment, the total cavitation dose was calculated as the cumulative sum of the cavitation dose for every pulse using MATLAB (Mathworks Inc., Natick, MA, USA).

$$SCD = \Sigma_{p=0\to100}SCD_p$$

$$ICD = \Sigma_{p=0\to100}ICD_p$$

Compared to stable cavitation, the frequency range to evaluate broadband emissions varies across different studies. To further compare the results acquired with different feedback algorithms, we re-analyzed the data based on another study where 1.15 ± 0.0125 MHz was defined for broadband emissions [35]. Finally, to achieve a more comprehensive evaluation, we applied a sliding 500 Hz window to the entire frequency spectrum (0-2 MHz) with a step size of 500 Hz and correlates the derived *ICD* value with the safety results.

To evaluate the prediction performance, we set up a specific threshold for positive labels and then calculated the ACC. For the efficacy prediction, we collected the *SCD* for all the failed experiments and used the maximum value as positive threshold. Then we compared the *SCD* for all successful experiments to this threshold. The predicting results were verified with the actual results of that experiment and the prediction ACC was then calculated by dividing the total number of correct predictions by the total number of experiments. For the safety prediction, we collected the *ICD* for all experiments showed hemorrhage and use the minimum value as positive threshold. The *ICD* of other experiments were then compared with this threshold. The predicting results were verified with histology evaluations and the prediction ACC was calculated by dividing the total number of correct predictions by the total number of samples.

Statistical analysis

One-way ANOVA was performed to evaluate the ACC/MCC performance among different training datasets and three different methods. Tukey adjustment for multiple comparison was used. Two-sample two-tailed t-test was used to evaluate the ACC/MCC performance between nanobubble and Definity. A p-value less than 0.05 was considered statistically significant. All analyses were performed in OriginPro, Version 2022 (OriginLab Corporation, Northampton, MA, USA).

**Results**

FUS-induced BBB opening outcomes in rats

Figure 3 shows the white light photo of perfused brain slices across the ultrasound beam at the level of the focus. Successful BBB opening was indicated by the localized Evans blue dye leakage, as shown in Figure 3A. In some cases, BBB opening was not successful as no dye leakage was observed (Figure 3B). Occasionally the BBB was successfully opened but there was tissue damage observed at the target (with hemorrhage, Figure 3C). The FUS-induced BBB opening outcomes are summarized in Table 2.

Machine learning-based efficacy prediction

*Prediction performance comparison of mSVDD model trained with different frequency bands*

Table 3 presents the mean and standard deviation of various evaluation metrics, including ACC, recall, precision, F1-score, and MCC, over twenty trials. These metrics were derived based on the mSVDD model, which was trained using labeled nanobubble data with frequency bands of 0-2 MHz and 0.7-0.8 MHz, respectively. As shown in Figure 4A, the ACC and MCC obtained from the overall bubble response (0-2 MHz) were significantly higher than those obtained from the ultra-harmonic bubble response (0.7-0.8 MHz) (ACC: $85.0 \pm 16.6\%$ vs $76.0 \pm 8.0\%$, p = 0.04, MCC: $0.7 \pm 0.3$ vs $0.5 \pm 0.2$, p = 0.03). As a result, we chose to use the overall frequency response of stimulated nanobubbles to train the models in the subsequent BBB opening outcome predictions.

*Effect of various models in prediction performance*

The mSVDD, iForest and MCD models were trained with the positive nanobubble data, and their performance was evaluated on the test dataset. As shown in Figure 4B, the trained mSVDD model outperformed the other two models in terms of both ACC and MCC with statistical significance ($p < 0.05$). The ACC and MCC for trained mSVDD model were $85.0 \pm 16.6\%$ and $0.7 \pm 0.3$, respectively, which are significantly higher than iForest model (ACC: $66.0 \pm 12.8\%$, MCC: $0.3 \pm 0.3$) and MCD model (ACC: $46.0 \pm 15.6\%$, MCC: $-0.1 \pm 0.3$). The detailed comparison of the performance of the three models is summarized in Table 4.

*Robustness of the trained model across different bubble formulations*

To evaluate the robustness and compatibility of the trained model across different bubble formulations, the best performing mSVDD model was chosen and retrained with the 0-2 MHz nanobubble training dataset. Subsequently, the retrained model was tested with both the nanobubble and the Definity dataset. As depicted in Figure 4C, the ACC achieved for nanobubbles ($83.0 \pm 18.7\%$) has a similar performance to that of Definity microbubble ($76.3 \pm 3.8\%$). However, the MCC achieved for nanobubbles was significantly higher than that of Definity ($0.7 \pm 0.3$ vs $-0.1 \pm 0.0$, $p < 0.0001$).

Machine learning-based safety prediction

*Prediction performance comparison of mSVDD model trained with different frequency bands*

The mean and standard deviation of various evaluation metrics over twenty trials are summarized in Table 5. As shown in Figure 5A, the mSVDD model had a significantly higher MCC when trained with the overall frequency response of stimulated bubbles (0-2 MHz) compared to the ultra-harmonic responses (0.7-0.8 MHz, $0.3 \pm 0.3$ vs $0.1 \pm 0.2$, $p = 0.04$). Similar performance was observed in terms of ACC (0-2 MHz: $62.5 \pm 12.8\%$, 0.7-0.8 MHz: $55.0 \pm 10.7\%$). Consequently, we utilized the overall response of stimulated nanobubbles for training in the subsequent safety prediction study.

*Effect of various models in prediction performance*

The performance of mSVDD model (ACC: $62.5 \pm 12.8\%$, MCC: $0.3 \pm 0.3$) and iForest model (ACC: $55.0 \pm 7.6\%$, MCC: $0.1 \pm 0.2$) were significantly better than the MCD model (ACC: $29.2 \pm 13.8\%$, MCC: $-0.4 \pm 0.3$), with details shown in Table 6 and Figure 5B.

*Robustness of the trained model across different bubble formulations*

The mSVDD model was selected and retrained on the 0-2 MHz nanobubble training data to evaluate the robustness and compatibility of the trained model across different bubble formulations. To assess the performance of the retrained model, the nanobubble testing dataset and the Definity dataset were utilized for evaluation of ACC/MCC metrics in twenty repeated trials. The ACC and MCC for nanobubbles were $62.5 \pm 12.8\%$ and $0.3 \pm 0.3$, which are significantly higher than Definity microbubbles (ACC: $36 \pm 8.2\%$, MCC: $-0.1 \pm 0.3$), as depicted in Figure 5C.

Conventional cavitation-based prediction for opening efficacy and safety

The cavitation doses (*ICD* and *SCD*) were calculated for the nanobubble experiments as shown in Figure 6. Figure 6A shows the *SCD* calculated for 35 experiments. Red and blue dots represent failed and successful experiments, respectively, and the red dashed line indicates the derived positive threshold. Based on this result the prediction ACC was 80.0%. For the safety prediction, there is no significant difference observed between the *ICD* of the experiments which BBB were safely opened and that of the experiments with hemorrhage (Figure 6B). The ACC of the prediction was 34.3%. The *ICD* recalculated at 1.15 MHz with a bandwidth of 25 kHz showed a prediction ACC of 40% (Figure 6C). To find the frequency range for *ICD* calculation that can best correlates with the safety results, we ran a comprehensive sliding window test with a bandwidth of 500 Hz and step size of 500 Hz. The analysis reveals that the highest accurate prediction was achieved at 0.236 MHz. Therefore, the *ICD* was recalculated at 0.236 MHz with a bandwidth of 25 kHz. The prediction ACC was improved to 57.1% (Figure 6D).

**Discussion**

Recently, researchers have increasingly proposed using acoustic feedback signals in specific frequency bands to predict the outcomes of FUS-induced BBB opening [33, 34] in addition to the two gold standards: contrast-enhanced MR imaging and histology evaluation [8, 36-38]. Despite the effectiveness of these cavitation-based methods, the algorithms are arbitrary and not very reliable across different experimental setups. With the vigorous development of machine learning methods in various fields [22, 23, 39, 40], we designed this study attempting to predict the FUS-induced BBB opening outcomes using machine learning with the acoustic signals collected during the treatments. Predictions of both the efficacy and safety of BBB opening were evaluated.

As summarized in Table 2 and shown in Figure 3, three outcomes of FUS BBB opening were achieved, including: successful opening, failed opening, and opening with hemorrhage. Because a previously developed real-time acoustic feedback control system [9] was used in this study, we had only 2 failed opening cases and 3 cases with hemorrhage out of 35 experiments in the nanobubble-assisted FUS-BBB opening. This resulted in an imbalanced dataset. To predict the BBB opening outcomes with this dataset, we first applied a machine learning model that is suitable for one-class classification [41]. The Support Vector Data Description (SVDD) [24] model with the radial basis function (RBF) kernel modified with a DTL method proposed by Xiao et al. [26] was investigated.

The best way to train the model was evaluated first. Our results indicate that the acoustic data within the overall frequency range should be considered in machine learning-based predictions, which is different from conventional cavitation-based methods, which usually consider certain frequency bands. The results in Figure 4A and Figure 5A demonstrated that 0-2 MHz trained model had a better performance compared to 0.7-0.8 MHz trained model in both opening efficacy and safety prediction. When using 0-2 MHz data to train the model, three primary metrics (ACC, Recall, and Precision) were greater than 80% (Table 3 and Figure 4A), indicating a robust and accurate BBB opening efficacy prediction. In the safety prediction, although the ACC did not show a significant difference, the MCC score obtained with 0-2 MHz data was significantly higher (Table 5 and Figure 5A). As widely accepted, ultra-harmonics (i.e., 0.7-0.8 MHz in this work) are more specifically related to the bubble activities under FUS exposures and are often used to indicate the BBB opening efficacy in conventional cavitation-based methods. However, in this study, they did not show a superior performance. One possible explanation is that machine learning methods can extract more information and features beyond our current existing knowledge.

Besides the mSVDD model, two commonly used anomaly detection methods: iForest model and MCD model were also evaluated and compared with the mSVDD model. In the BBB opening efficacy prediction, as shown in Figure 4B and Table 4, the mSVDD model showed significantly better performance in both ACC and MCC than the iForest and MCD models. And the iForest model showed significantly better ACC and MCC results than those of the MCD model. In the safety prediction (Figure 5B and Table 6), the ACC and MCC performance of mSVDD and iForest are significantly better than that of the MCD model. Possible reasons for these results may include: 1) mSVDD is a supervised algorithm while iForest and MCD algorithms are unsupervised, so that the mSVDD is able to delineate the decision boundary between normal and anomalous samples more accurately; 2) mSVDD uses Gaussian kernels to fit the nonlinear data, while MCD and iForest can only handle the linear data. Hence, the mSVDD model may outperform other models on nonlinear datasets in practical applications; and 3) The mSVDD and iForest models perform well in high-dimensional data, which is our case. Due to the computational and storage requirements for the MCD model, the PCA was used to reduce the dimensionality, which may lead to less features, therefore less satisfied performance of the MCD model compared to the other two models.

Currently, different bubble formulations are used for FUS-BBB opening, such as: Definity, Optison, SonoVue, and other custom-made bubbles in laboratories. To test the robustness and compatibility of the machine learning model, we trained our model with datasets acquired with nanobubbles and tested its performance on Definity. Our findings indicate that the machine learning model trained with one bubble type cannot be directly applied on different bubble formulations, given their significantly different performance (Figure 4C and Figure 5C). Interestingly, as shown in Figure 4C, despite the ACC value on Definity in the efficacy prediction was high (76.3 ± 3.8%), the MCC value was negative (-0.1 ± 0.0). To further investigate the reason behind the negative MCC value and high ACC value, we examined the confusion matrix and found that the low TN value is the reason. Moreover, the MCC value on the Definity in the safety prediction was close to zero (Figure 5C), indicating that the model's ability to predict the safety of the FUS BBB opening on Definity was random. Because different bubble formulations vary in size, distribution, and other properties, it is challenging to transfer features extracted from one bubble type to another. There are two directions that can be pursued in future: 1) Specific models are trained for each bubble formulation; and 2) Train the machine learning model with a larger training dataset comprised of multiple bubble formulations, so that it could extract common features among different bubbles to generate a valid model for a more generic prediction.

As mentioned above, researchers use stable and inertial cavitation as indicators for BBB opening outcomes prediction, as a convenient method in compensation for the MRI and histology assessment. Conventionally, *SCD* is used to predict opening efficacy, and *ICD* is used as an indicator for safety. The detection of *SCD* is often linked to successful BBB opening, while the detection of *ICD* indicates hemorrhage or tissue damage. In this study, *SCD* worked with a high prediction ACC as expected for predicting the opening efficacy (80.0%, Figure 6A), slightly lower than the best results of machine learning-based prediction (85.0%). However, *ICD* did not correlate well with the safety results. As shown in Figure 6B-C, the prediction ACC was between 34%-40%. With our setup, the best frequency band for *ICD* dose calculation was found to be $236 \pm 25$ kHz for nanobubbles and it achieved the highest prediction ACC of 57.1% (Figure 6D). However, the performance was still not as good as the one achieved with mSVDD model (62.5%). This suggests that machine learning-based FUS-BBB opening outcome prediction might be able to overcome the current limitations of cavitation-based methods and to even further improve the performance. More studies are warranted to make comparisons between the two methods with a larger dataset and different experimental settings.

There are several limitations in this study. Firstly, due to experimental costs, animal welfare, and lack of publicly available datasets, the sample size for the model training is limited. For testing the compatibility with different bubble formulations, ideally more than two bubble formulations should be tested and included in the training dataset, which is one of the primary tasks we are planning for future studies. Secondly, only three models were evaluated in this study, other machine learning or deep learning models may be more suitable for this task and deserves more investigation.

**Conclusion**

Our results indicate that it is feasible to reliably predict the FUS-induced BBB opening outcomes using machine learning with the acoustic signals generated from stimulated nanobubbles. Among the three machine learning models tested in this study, the mSVDD model outperforms the other two: iForest and MCD models. Different from conventional cavitation-based methods, it is not necessary to use acoustic data within a certain frequency range(s) in both efficacy and safety prediction with machine learning. In addition, our findings indicate that the model trained with one bubble type cannot be applied directly on different bubble formulations. Lastly, machine learning-based methods produced comparable or even better prediction results compared to conventional cavitation-based methods. Further studies are required to improve the performance of the machine learning-based predictions.

**Abbreviations**

ACC: accuracy; AUC: area under the curve; BBB: blood-brain barrier; FUS: focused ultrasound; FFT: fast Fourier transform; ICD: inertial cavitation dose; iForest: isolation forest; MCC: Matthew's correlation coefficient; MCD: minimum covariance determinant; MD: Mahalanobis distance; MR: magnetic resonance; MRI: magnetic resonance imaging; MR-HIFU: magnetic resonance-guided high-intensity focused ultrasound; mSVDD: modified support vector data description; PCA: principal component analysis; RBC: red blood cell; RBF: radial basis function; SCD: stable cavitation dose; T1w MRI: T1-weighted magnetic resonance imaging.

## Acknowledgements

This work was supported in part by funding from the Natural Science Foundation of Shanghai (23ZR1442000, BC) and the start-up grant from ShanghaiTech University (2021F0209-000-09, BC). The authors acknowledge the use of BioRender that is used to create schematic Figure 1.

## Competing interests

The authors have declared that no competing interest exists.

**Tables**

Table 1. The confusion matrix

|  |  | Actual label | |
|---|---|---|---|
|  |  | Positive | Negative |
| **Prediction label** | True | True Positive (TP) | True Negative (TN) |
|  | False | False Positive (FP) | False Negative (FN) |

Table 2. Summary of the BBB opening outcomes

|  | Successful | | Failed |
|---|---|---|---|
|  | No hemorrhage | With hemorrhage |  |
| **Nanobubble** | 30 | 3 | 2 |
| **Definity** | 2 | 3 | 1 |

Table 3. mSVDD performance comparison across training datasets in different frequency range for efficacy prediction

|  | **ACC (%)** | **Recall (%)** | **Precision (%)** | **F1-Score (%)** | **MCC** |
|---|---|---|---|---|---|
| **0-2 MHz** | 85.0 ± 16.6 | 81.7 ± 19.6 | 82.9 ± 31.2 | 77.0 ± 29.8 | 0.7 ± 0.3 |
| **0.7-0.8 MHz** | 76.0 ± 8.0 | 93.3 ± 13.3 | 65.8 ± 22.2 | 73.3 ± 25.6 | 0.5 ± 0.2 |

Table 4. Performance comparison of mSVDD, iForest, and MCD in BBB opening efficacy prediction

|  | **ACC (%)** | **Recall (%)** | **Precision (%)** | **F1-Score (%)** | **MCC** |
|---|---|---|---|---|---|
| **mSVDD** | 85.0 ± 16.6 | 81.7 ± 19.6 | 82.9 ± 31.2 | 77.0 ± 29.8 | 0.7 ± 0.3 |
| **iForest** | 66.0 ± 12.8 | 76.7 ± 21.3 | 68.3 ± 7.3 | 71.6 ± 13.8 | 0.3 ± 0.3 |
| **MCD** | 46.0 ± 15.6 | 43.3 ± 26.0 | 49.6 ± 22.4 | 45.6 ± 23.7 | -0.1 ± 0.3 |

Table 5. mSVDD performance comparison across training datasets in different frequency range for safety prediction

|  | **ACC (%)** | **Recall (%)** | **Precision (%)** | **F1-Score (%)** | **MCC** |
|---|---|---|---|---|---|
| **0-2 MHz** | 62.5 ± 12.8 | 25.0 ± 25.5 | 60.0 ± 49.0 | 34.0 ± 30.2 | 0.3 ± 0.3 |
| **0.7-0.8 MHz** | 55.0 ± 10.7 | 18.3 ± 22.3 | 35.0 ± 42.5 | 23.0 ± 27.0 | 0.1 ± 0.2 |

Table 6. Performance comparison across different models for safety prediction

|  | **ACC (%)** | **Recall (%)** | **Precision (%)** | **F1-Score (%)** | **MCC** |
|---|---|---|---|---|---|
| **mSVDD** | 62.5 ± 12.8 | 25.0 ± 25.5 | 60.0 ± 49.0 | 34.0 ± 30.2 | 0.3 ± 0.3 |
| **iForest** | 55.0 ± 7.6 | 10.0 ± 15.3 | 30.0 ± 45.8 | 15.0 ± 22.9 | 0.1 ± 0.2 |
| **MCD** | 29.2 ± 13.8 | 36.7 ± 23.3 | 29.3 ± 16.2 | 32.3 ± 18.7 | -0.4 ± 0.3 |

# Figures

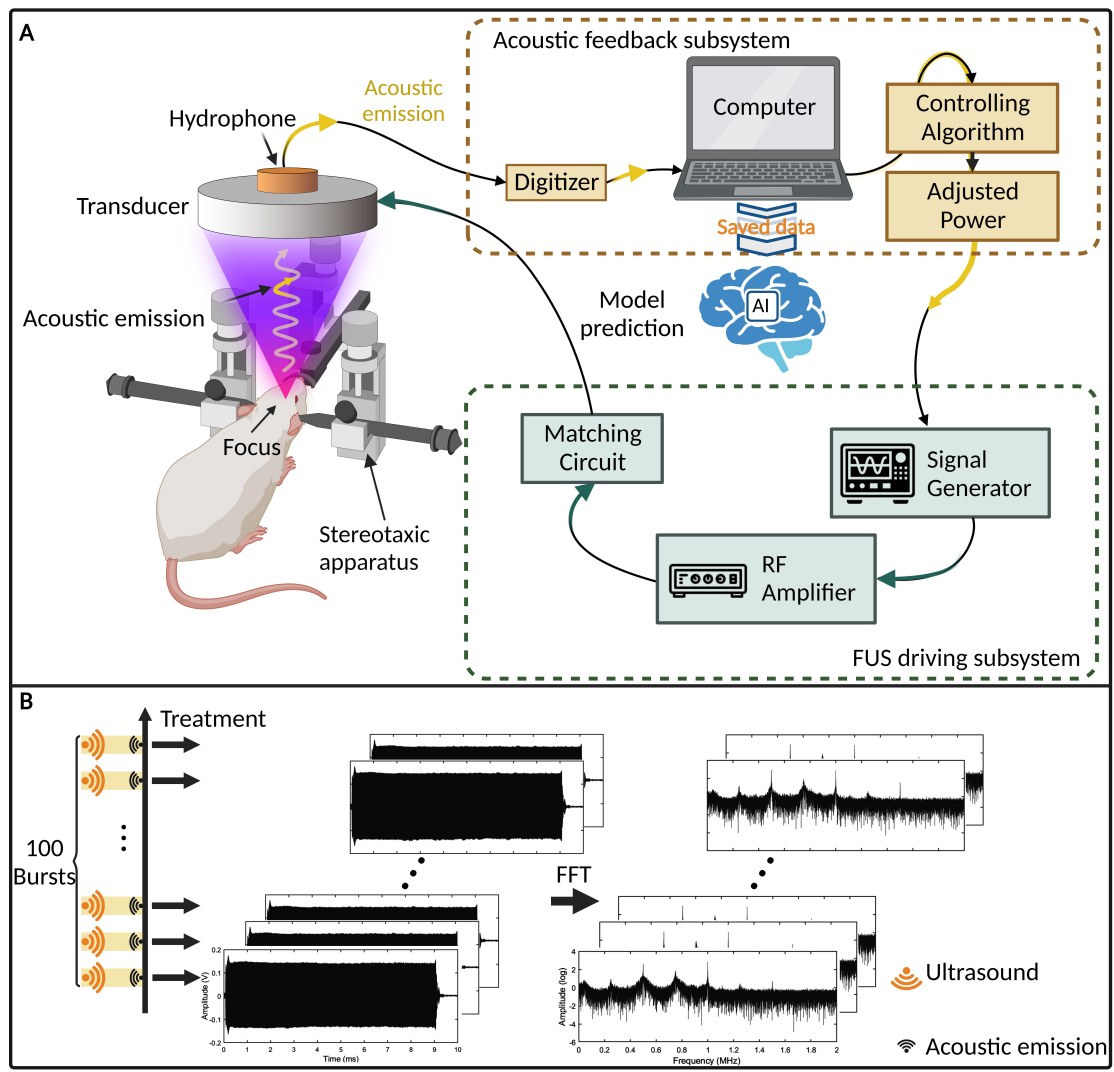

**Figure 1. Acoustic feedback-controlled FUS-induced BBB opening. A)** The BBB opening system is comprised of a FUS driving subsystem and an acoustic feedback subsystem. The FUS driving subsystem includes a signal generator, RF amplifier, matching circuit and a focused ultrasound transducer. The acoustic feedback subsystem includes a hydrophone, digitizer, built-in controlling algorithm, and a PC; **B)** The treatment protocol and acoustic signals from stimulated nanobubbles.

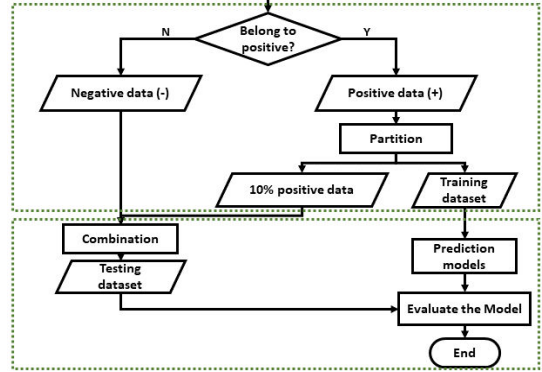

**Figure 2. The flowchart of the prediction algorithm.**

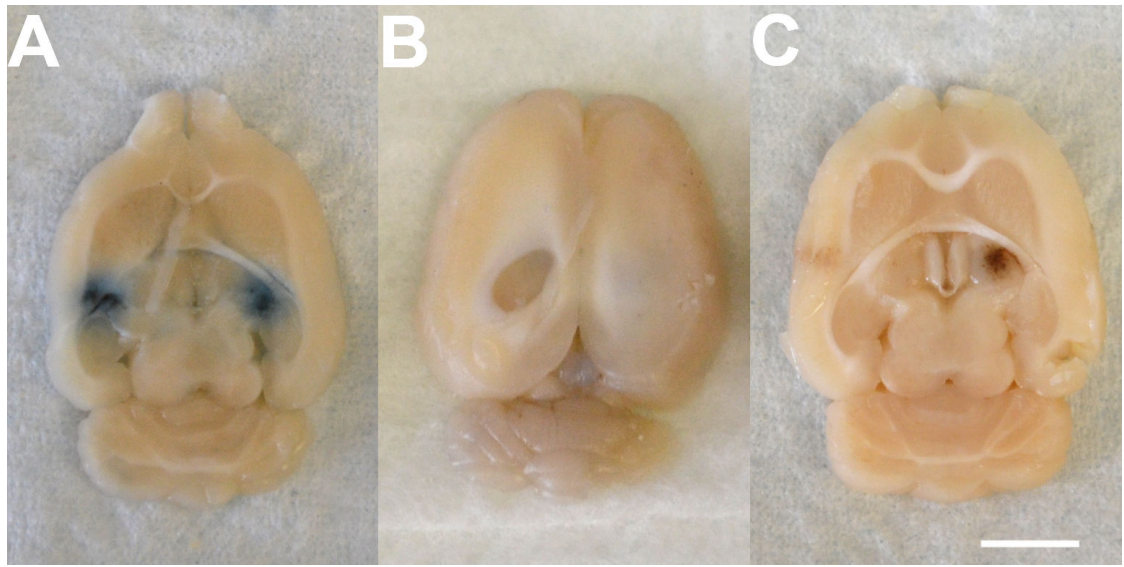

**Figure 3. Gross-tissue photos of the brain slice with Evans blue dye leakage indicating the opening outcomes. A)** Successful BBB opening; **B)** Failed BBB opening (no dye leakage); **C)** BBB opening with hemorrhage. BBB: blood-brain barrier. The white scale bar indicates 5 mm.

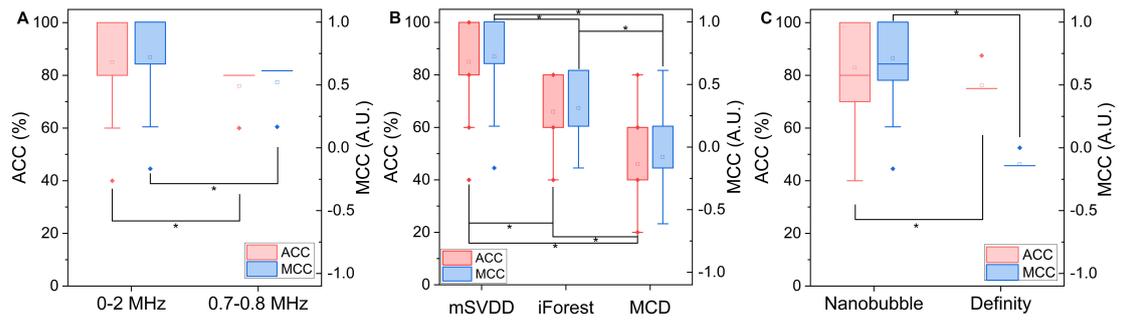

**Figure 4. Machine learning-based efficacy prediction. A)** Performance of mSVDD model trained with different frequency data of nanobubbles; **B)** Performance comparison across different machine learning models; **C)** mSVDD model-based prediction performance comparison across different bubble formulations. *: $p < 0.05$. Error bar: Standard deviation (n=20). ACC: Accuracy (left y-axis, red); MCC: Matthew Correlation Coefficient (right y-axis, blue).

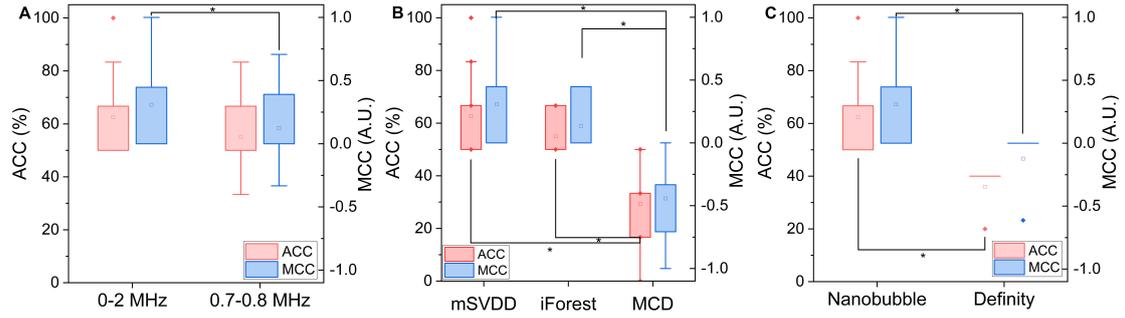

**Figure 5. Machine learning-based safety prediction. A)** Performance of mSVDD model trained with different frequency data of nanobubbles; **B)** Performance of different machine learning models on BBB opening safety prediction; **C)** mSVDD model-based safety prediction performance comparison across different bubble formulations. *: $p < 0.05$. Error bar: Standard deviation (n=20). ACC: Accuracy (left y-axis, red); MCC: Matthew Correlation Coefficient (right y-axis, blue).

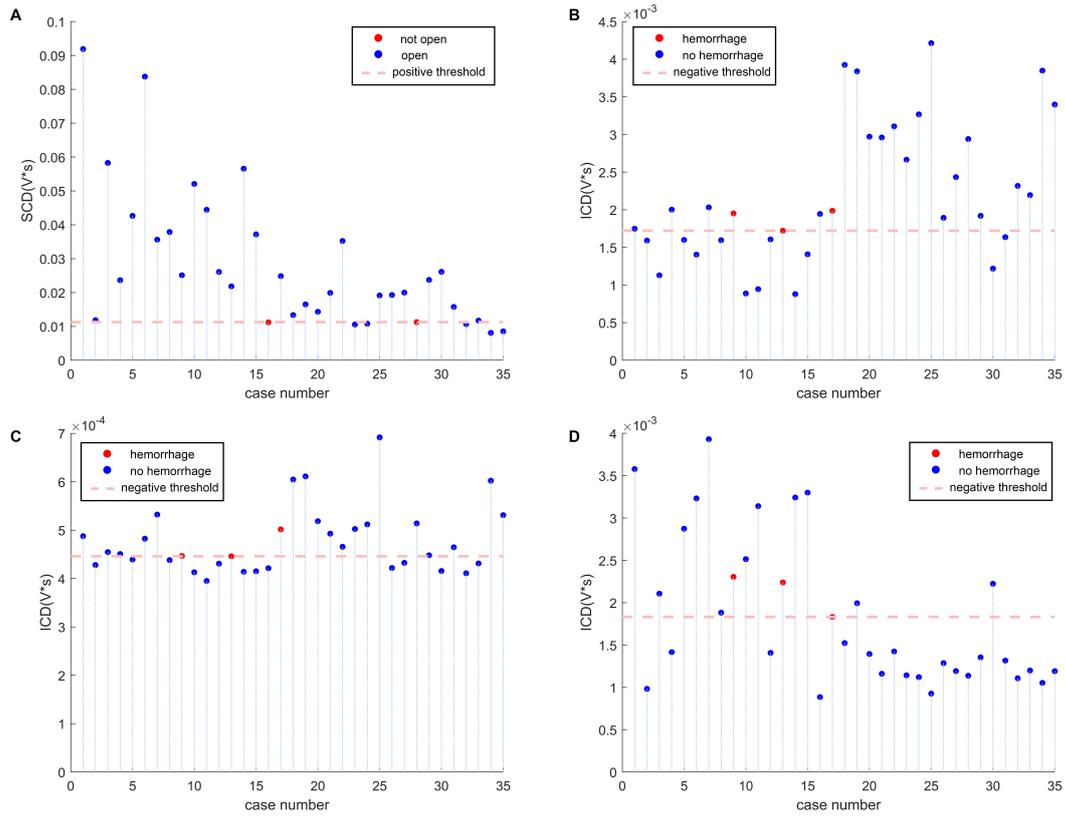

**Figure 6. Stable and inertial cavitation dose calculated from in-vivo BBB opening experiments. A)** The $SCD$ of 35 experiments injected with nanobubbles. Based on the threshold (red dashed line) the prediction ACC was 80.0% for the opening efficacy; **B)** The $ICD$ of 35 experiments injected with nanobubbles. Harmonics (harmonic frequency ± 180 kHz) and ultra-harmonics (ultra-harmonic frequency ± 50 kHz) were excluded when calculate the $ICD$. Based on the threshold (red dashed line) the prediction ACC was 34.3% for safety; **C)** The recalculated $ICD$ (at 1.15 MHz ± 12.5 kHz) of 35 experiments with nanobubbles, showing a prediction ACC of 40%; **D)** The recalculated $ICD$ (at 0.236 MHz ± 12.5 kHz) of 35 experiments with nanobubbles, the prediction ACC was improved to 57.1%. ACC: Accuracy.